# Radiometric Actuators for Spacecraft Attitude Control

**Ravi Teja Nallapu**
**Space and Terrestrial Robotic Exploration Laboratory**
**Arizona State University**
**781 E. Terrace Mall, Tempe, AZ**
**rnallapu@asu.edu**

**Amit Tallapragada**
**Space and Terrestrial Robotic Exploration Laboratory**
**Arizona State University**
**781 E. Terrace Mall, Tempe, AZ**
**atallapr@asu.edu**

**Jekan Thangavelautham**
**Space and Terrestrial Robotic Exploration Laboratory**
**Arizona State University**
**781 E. Terrace Mall, Tempe, AZ**
**jekan@asu.edu**

*Abstract*—**CubeSats and small satellites are emerging as low-cost tools to perform astronomy, exoplanet searches and earth observation. These satellites can be dedicated to pointing at targets for weeks or months at a time. This is typically not possible on larger missions where usage is shared. Current satellites use reaction wheels and where possible magneto-torquers to control attitude. However, these actuators can induce jitter due to various sources. In this work, we introduce a new class of actuators that exploit radiometric forces induced by gasses on surface with a thermal gradient. Our work shows that a CubeSat or small spacecraft mounted with radiometric actuators can achieve precise pointing of few arc-seconds or less and avoid the jitter problem. The actuator is entirely solid-state, containing no moving mechanical components. This ensures high-reliability and long-life in space. A preliminary design for these actuators is proposed, followed by feasibility analysis of the actuator performance.**

## TABLE OF CONTENTS

## 1. INTRODUCTION

CubeSats and small satellites are emerging as low-cost tools to perform astronomy, exoplanet searches and for earth observation. These satellites can be dedicated to pointing at a target for weeks or months at a time. This is typically not possible on larger missions where usage is shared. The problem of designing a precise attitude control system for CubeSats and small satellites is very challenging: Not only are the choice of actuators limited to reaction wheels and magneto-torquers, but the existing actuators can also induce jitter on the spacecraft due to moving mechanical parts. In addition, such a spacecraft may contain cryo-pumps and servos that introduce additional vibrations. Therefore, an actuator to control the attitude and dampen spacecraft vibrations simultaneously is needed. Techniques to overcome reaction wheel jitter exist, they require additional devices such as piezo-electric actuators to compensate for jitter and stabilize a payload [14].

With this in mind, a new device to control the orientation of a spacecraft is presented in this paper and exploits radiometric forces. To understand how this actuator works, let's look at the Crooke's Radiometer [1, 2] which consists of 4 thin plates, called vanes, each colored black and white on its opposite surfaces. These vanes are suspended from a spindle such that opposite colors faces each other. This setup is now placed in a low-pressure vessel containing trace gasses of nitrogen or argon (Figure 1). When this device is exposed to a heat source, the different colored vane surfaces acquire different temperatures, i.e., the black surface acquires a higher temperature than the white surface. If this temperature difference happens within a 'thin' gradient, then the gas particles at the black surface strike it with higher momentum than those at the white surface causing the spindle to spin toward the white plates.

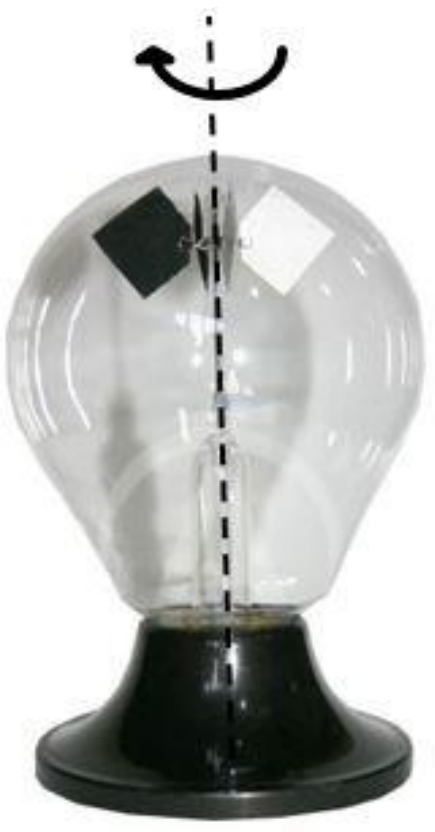

**Figure 1. Crookes radiometer**

We propose to implement this radiometer as an attitude control actuator on a CubeSat by installing 8 thin vanes of different temperatures placed under low pressure. By controlling the temperature of the vanes, we can generate the torque needed to precisely point or spin a spacecraft. These chambers contain trace amounts of a stable gas like Nitrogen or Argon. In this setup, the temperature gradient across the vanes causes the gas molecules to strike the vanes differently and thus inducing this radiometric force. This paper talks about design and feasibility analysis of one such actuator that can enable precision pointing within an arc-second and achieve spin of 1 RPM or more.

Continuously spinning a CubeSat sized spacecraft has important applications, particularly for generating artificial gravity [20-21]. AOSAT 1 [19, 22] is a CubeSat centrifuge science laboratory under development and can potentially benefit from this technology for attitude control. Use of radiometric actuators could replace expensive reaction wheels.

The rest of this paper is organized as follows: We present a system overview of the proposed actuators in Section 2. This is followed by brief discussion of the theory behind the operation of these actuators in Section 3. Section 4 presents detailed description of the actuator design. In section 5 we present the results of simulations performed on the theoretical design mentioned in the previous section. This is followed by a discussion and conclusions in Section 6 and 7 respectively.

## 2. SYSTEM OVERVIEW

The proposed actuator uses the principle of radiometric forces discussed in Section 1. Figure 2 shows a conceptual actuator.

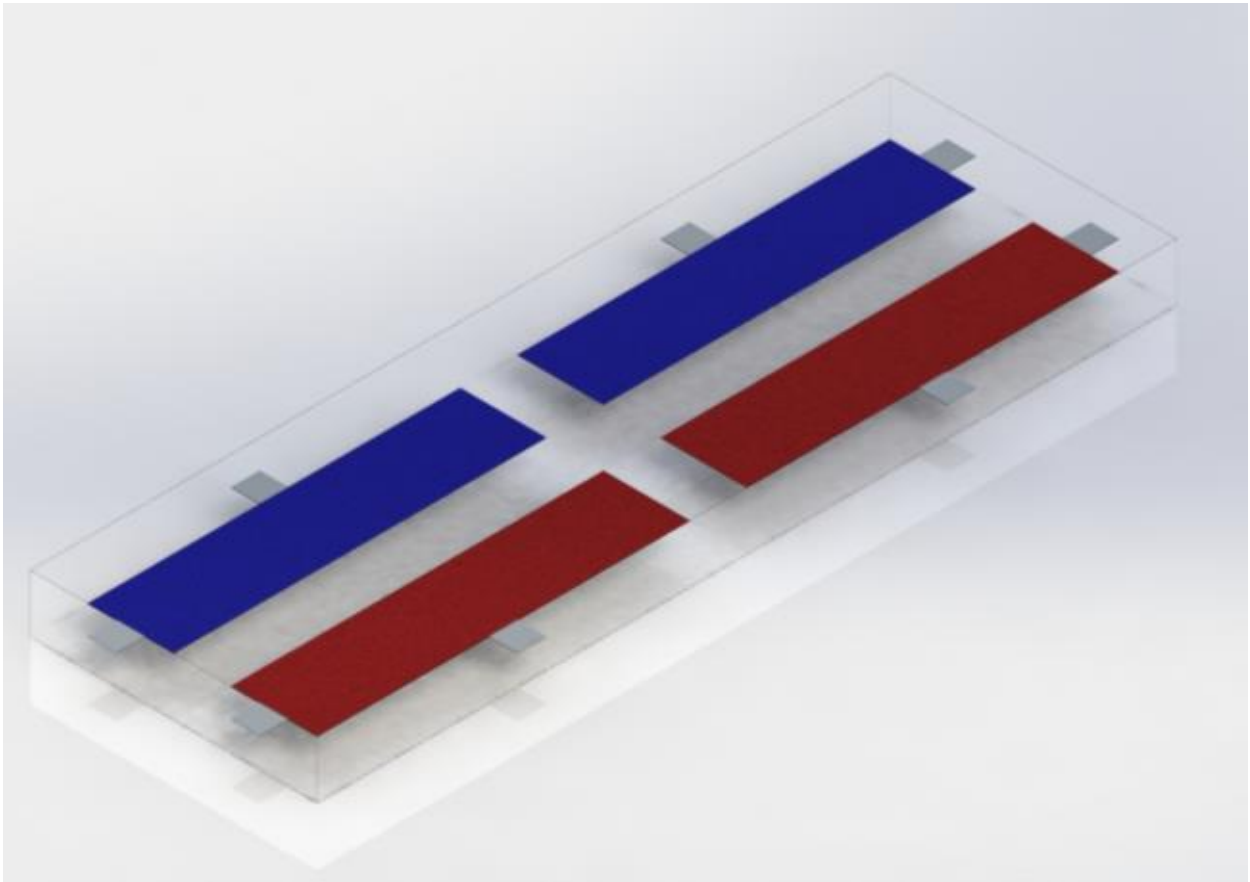

**Figure 2. Concept radiometric actuator**

This actuator is designed to fit on each face of a 3U CubeSat. It contains a partial vacuum chamber filled with trace amounts of a stable gas such as Argon or Nitrogen. Eight vanes are arranged in the form of a two 2 × 2 matrix. These matrices are mounted to the chamber as shown in Figure 2. The temperature of each vanes can be controlled using Thermo-Electric Devices (TEDs).

The radiometric forces are only effective when the temperature difference exists in a small gradient, so by spacing the 2 vane matrices ‘close’—on the order of the mean free path of the gas [3], this will be discussed in Section 4. Thus, by controlling the temperature of the vanes, the direction of the gradient can be changed, thus changing the direction of gas flow. This induces forces on the vanes, thus causing the whole setup to rotate in space.

Figure 3 presents a diagram of the proposed actuators. The actuators require a micro-controller that implements a feed-back control system that takes in measured angle and angular velocity from an IMU, together with temperatures readings from the vanes. The output from the controller consists of a series of commands to the heating and cooling elements of the vanes. These sensory inputs are then use to precisely control the heating and cooling elements.

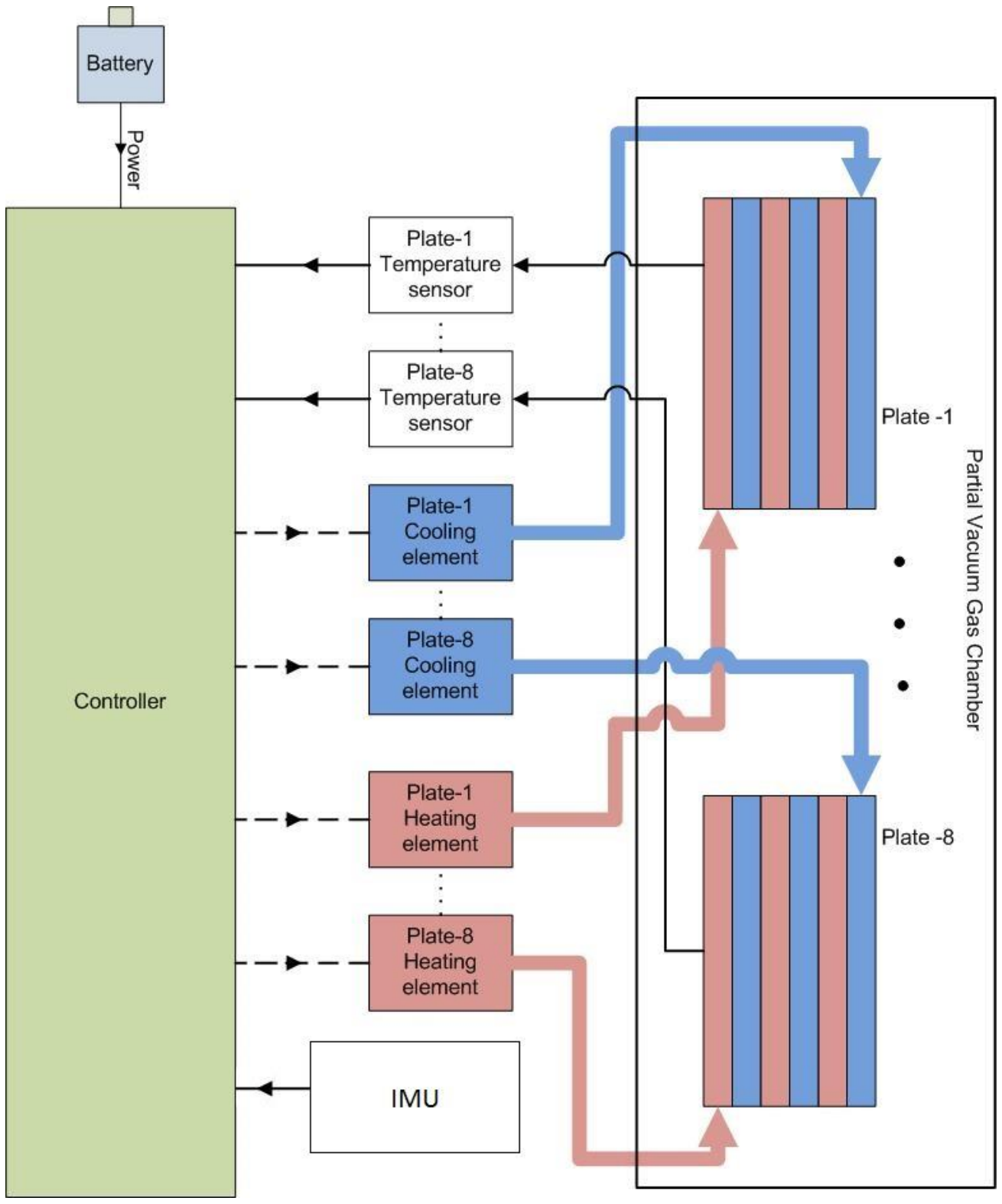

**Figure 3. Schematic of the radiometric actuator**

## 3. THEORY

Radiometric forces are caused by the interplay of two phenomena: The thermal creep forces or the shear force ($F_S$), and the normal force or the pressure force ($F_N$), as show in Figure 4. A brief introduction to these forces along with a summary of the models will be presented here.

Consider a hot-cold vane pair as shown in Figure 4. Assume the two vanes to be geometrically identical, and be made of same material, with a length, $l$, and width, $w$. Without loss of generality let one of the vanes be hotter than the other with a temperature $T_h$, and let the colder vane have a temperature $T_c$. Let the two vanes contribute an effective thickness, $d$, and finally let this setup be placed in a partial vacuum chamber with a gas having a mean free path, $\lambda$, and have a temperature, $T$. This is a modified setup than the one observed described by Sandurra [3], however the same reasoning has been applied by Selden and Galande [5], for a setup comparable to our radiometric actuator.

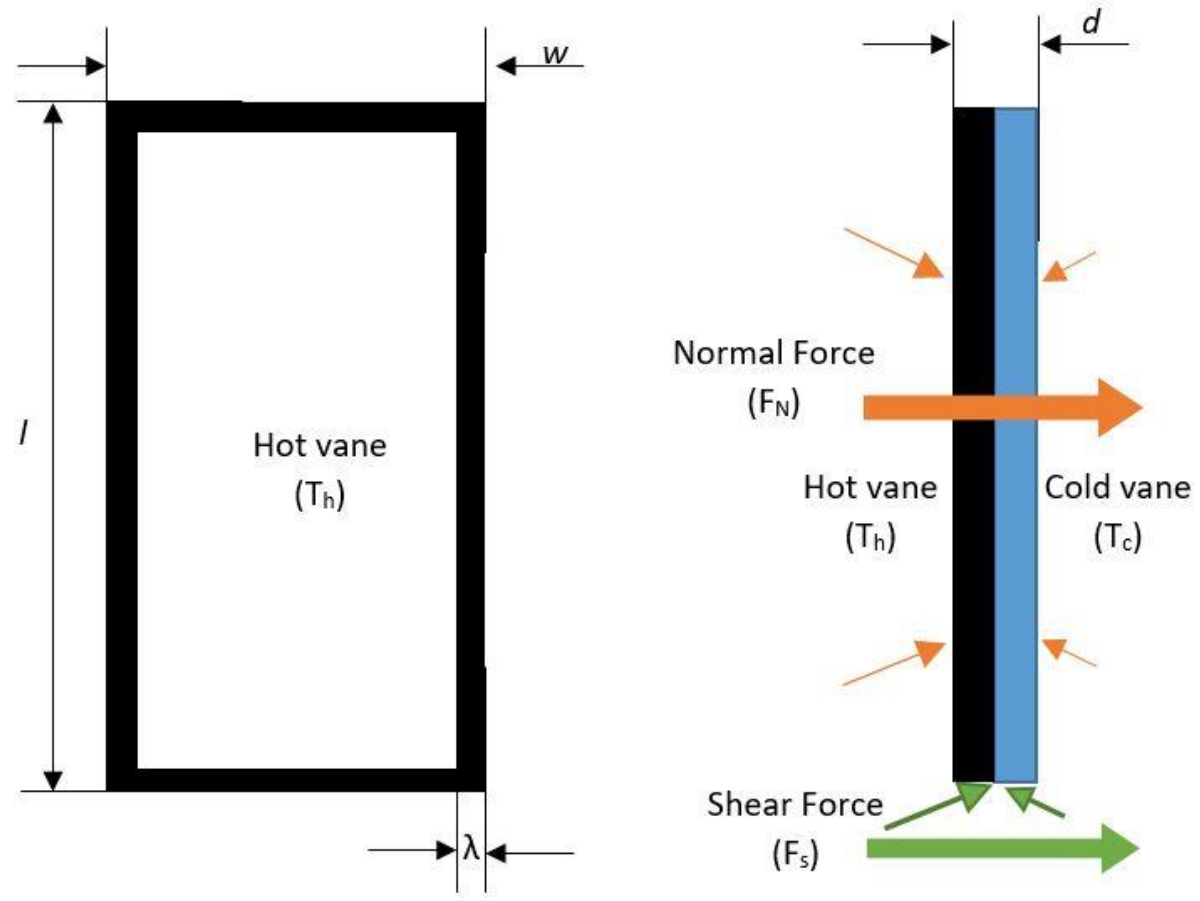

**Figure 4. Vanes creating the temperature gradient**

*Normal Force*

Since the gas is heated unequally on either side of the vanes, the molecules on one side strike the vane with higher velocities than those striking on the colder side. Because of this difference in the momentum deposition, a net force towards the cold side is generated. This is the normal force ($F_N$). This force was for a while thought to be the only cause for the Crooke's radiometer. However, it was argued and shown later that if this was the only cause, the radiometer would eventually slowdown, because equilibrium would eventually be reached.

*Shear Force*

The shear force or the thermal creep force is caused because of the molecules striking at the hot-cold interface along the vane thickness as shown in Figure 4. This force is an important cause for a steady-state spin of the radiometer. It is important here that the length of the gradient, in this case the combined thickness $d$, be comparable to mean free path $\lambda$ of the gas molecules considered. If this requirement is not met, the gas collisions dominate and the creep force vanishes.

Sandurra in [6] derived expressions for these two forces by solving the Boltzmann's equation with a BGK approximation [17-18], which is given by the following:

$$F_N = (2-\alpha)\frac{15}{32\sqrt{2}\,\pi}\frac{k}{\sigma^2}\frac{\Delta T}{d}(\lambda P) \qquad (1)$$

and,

$$F_s = \frac{15}{64\sqrt{2}\,\pi}\frac{k}{\sigma^2}\,\alpha\,\frac{\Delta T}{\lambda}\,(dP) \qquad (2)$$

Where $k$ is the Boltzmann's constant, and $\alpha$ is the accommodation coefficient of the vane surfaces which is the fraction of total number of collisions that are diffuse, $\sigma$ is the diameter of the gas molecules. $P$ is the perimeter of the vane surfaces and,

$$P = 2\,(l + w). \qquad (3)$$

If the vanes have a larger thickness than the length (t > l), the shear forces dominate, and if we have narrow vanes where (t< l), the normal force dominates [6]. Now if we define the Local Knudsen Number Kn [3,6] as:

$$Kn = \frac{\lambda}{d} \qquad (4)$$

and a constant $\varphi$ as:

$$\varphi = \frac{15}{32\sqrt{2}\pi}\frac{k}{\sigma^2} \qquad (5)$$

Then we can express the total force ($F_{rad}$) as the sum of these forces as:

$$\mathrm{F_{rad}} = \varphi\,\Delta T\,P\left[2 + \alpha\left(\frac{1}{2Kn} - 1\right)\right] \qquad (6)$$

Also, from the kinetic theory of gasses [17], the mean free path of a gas at pressure $P_{gas}$ given by:

$$\lambda = \frac{\mathrm{kT}}{\sqrt{2}\,P_{gas}\sigma} \qquad (7)$$

We refer the reader to Sandurra [6], for a complete understanding of the derivation of Equation (6). One comment that needs to be made here is that if $\alpha$=1, i.e., if all the collisions are diffuse, the radiometric force is maximized.

However, the magnitude of the shear force differs from that in (2) because of a reduction in effective area on which the force acts. The existence of multiple temperature gradients is shown in Figure 5. As shown here, at a hot-cold interface 4 gradients come into play. Gradients 1 and 2 and are caused by the temperature difference between ambient air and the hot vane. Gradient 4 produces the dominant force, and thus drives the mechanism. Gradient 2 produces an opposing force to that caused by Gradient 1. However, this can be modeled as a reduction in the main force rather than an extra force. Gradients 1 and 3 produce equal and opposite forces thus nullifying each other. However, they reduce the effective area as shown in Figure 5.

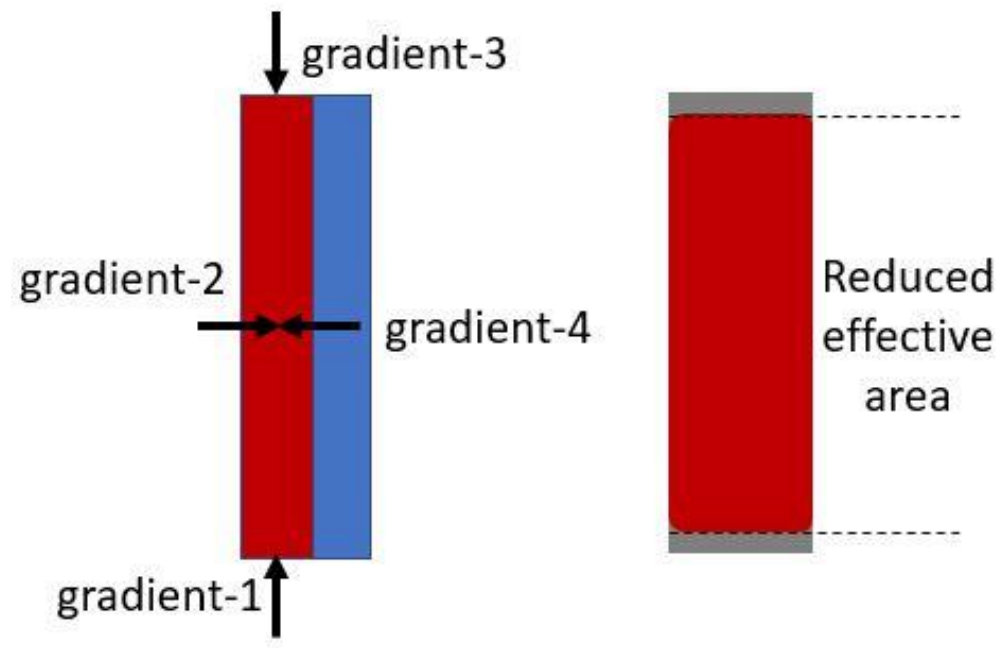

**Figure 5. Reduction in Shear Force**

Because of this modification, the net shear force is given by:

$$F'_s = \frac{\mathrm{x}15}{32\sqrt{2}\,\pi}\frac{k\Delta T\alpha}{\sigma^2}(\mathrm{l}' + w')\min\left(\frac{d}{W_{grad}}, 1\right) \qquad (8)$$

where $l'$ and $w'$ are the changed effective length and width given by:

$$l' = l - \beta\lambda\left(\frac{2-\alpha}{\alpha}\right) \quad (9)$$

and

$$w' = w - \beta\lambda\left(\frac{2-\alpha}{\alpha}\right) \quad (10)$$

The effective gradient width in the gas, $w_{grad}$, is given by:

$$w_{grad} = d + 2\lambda\left(\frac{2-\alpha}{\alpha}\right) \quad (11)$$

The parameters $x$ and $\beta$ are numerical correction factors that compensate for reduction in the shear force due to Gradient 2, and the reductions due to Gradients 1 and 3, respectively. These are found experimentally, and vary between 0 and 1.

Finally, the net radiometric force generated by the actuator shown in Figure 2 is given by:

$$F = N(F_N + F'_S). \quad (12)$$

Where $N$ is the total number of hot-cold vane pairs. This linear formulation of the force, is very helpful in gauging the performance of the actuator on the orders of magnitude.

## 4. DESIGN

The problem of designing an actuator can be divided into three major steps: (1) Geometrical Sizing (sizing the vanes and the box), (2) Material Selection (choosing the gas and vane material), and (3) Performance Evaluation (maximum forces and torques generated). This section describes the technical considerations that go into the actuator design for a 3U CubeSat:

*Geometrical*

The magnitude of force in Equation (6) is directly proportional to the vane perimeter, it is preferable to have it maximized; Therefore, the actuators can be mounted on the outer surface of the CubeSat as shown in Figure 5 and adhering to the CubeSat standards [15]. This means that the casing has a length of about 34 cm, and a width of 10 cm as shown in Figure 6.

Each vane is about $l$=13 cm in length and $w$=3 cm wide. A nominal thickness of 0.5 mm is considered, therefore the gradient thickness ($d$=1 mm).

The Knudsen number described in Equation (4) is an important parameter for characterizing this force [10]. If the Knudsen number is too small (Kn<0.001), then the equations described in Section 3 would not be valid and will have to be replaced by the continuum equations because the system is in the slip flow regime.

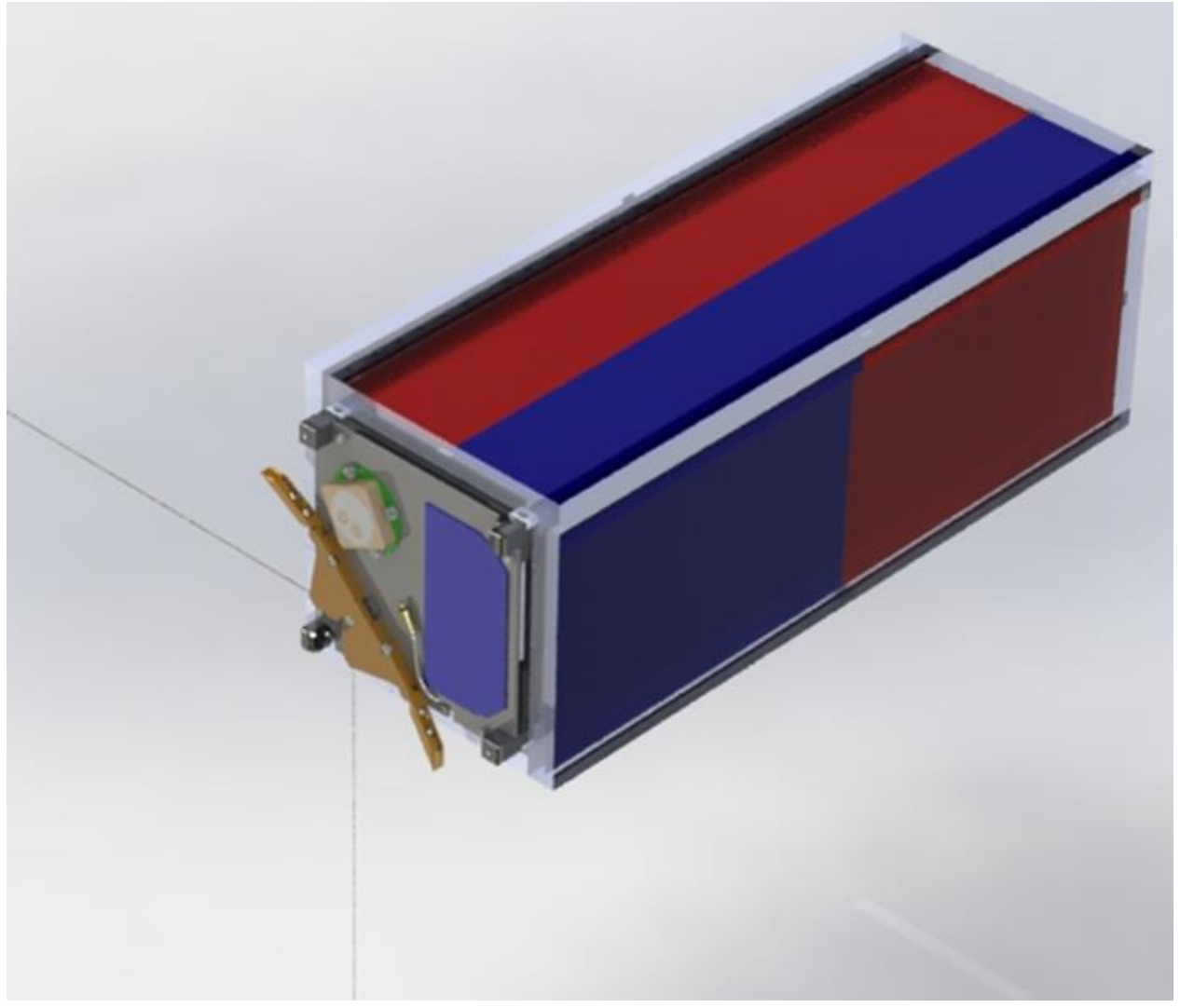

**Figure 6. 3U-CubeSat with radiometric actuators**

In the slip flow regime, viscous forces and other losses dominate, and the radiometric effect diminishes. In contrast, if the Knudsen number is high ($Kn$>10) [10-11], then the gas molecules cannot impart sufficient momentum for the motion, this regime is called the free molecular regime. Only when the Knudsen number is between the above two values, (0.001<$Kn$<10), we expect the radiometric force to occur. This regime of gas flows is called the transitional regime since both gas and fluid based laws can be applied [10-11]. The existence of a peak force in this domain was proved by [12]. With this in mind, the 2 vane matrices are separated by a distance equal to the mean free path of the gas ($d = \lambda$).

*Material*

The vane material strictly speaking has very little impact on the net radiometric force. The accommodation co-efficient is the only surface parameter, and as discussed earlier, the radiometric forces are maximum when $\alpha$=1.

The gas on the other hand has several considerations to be made. A heavier gas has larger diameter, so the force is reduced. Therefore, lighter gases produce more forces. However, Sexl [13] showed that the radiometric force is inversely proportional to the molecular degrees of freedom, which means monoatomic molecules are the most effective ones. Nitrogen however has been used in several radiometric studies in the past because it's easy to obtain and is inert. Helium is the ideal gas for generating radiometric forces, as it's light and inert. However, helium is hard to store for long durations.

The dependence of the mean free path on the gas pressure is presented in Figure 7, as predicted by Equation 7. Since our effective thickness is 1 mm, we note the pressures where the mean free path is 1 mm and this is presented in Table 1.

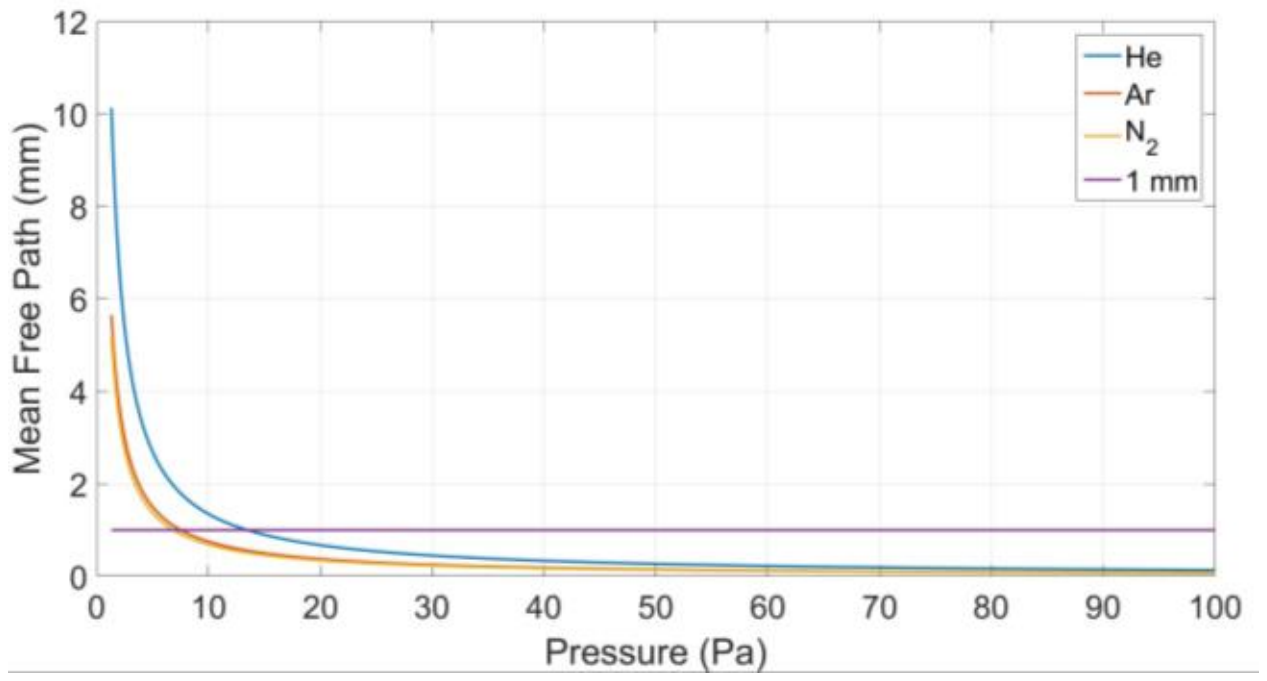

**Figure 7. Dependence of Mean free path on gas pressure**

*Performance*

At an Earth-Moon-Sun Lagrange point, Solar Radiation Pressure is the major cause of disturbance torques [14]. Figure 8 shows the variation of solar radiation pressure at 1 AU from the sun, where L-1 and L-2 points can be approximated. It can be seen in Figure 6, that the maximum solar torques can be expected to be 2.1E-9 N-m. So we can design an actuator that can compensate for this disturbance with 10 times its magnitude, i.e., 2.1E-8 N-m. Thus we need to design an actuator that can at least counter act this disturbance.

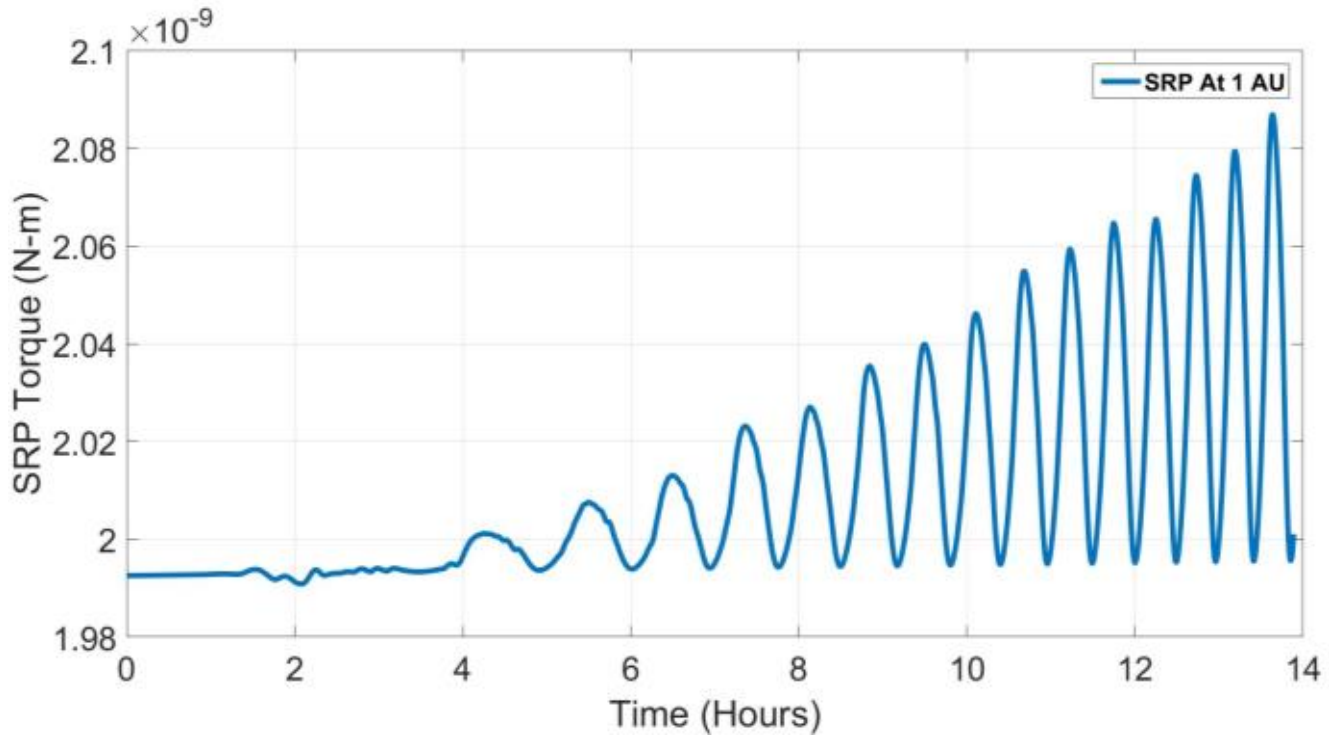

**Figure 8. SRP Disturbance at 1 AU**

## 5. RESULTS

A simulation was run to analyze the performance of the above-mentioned design with the parameters x=0.5 and β=0.5. The results are presented here:

*Magnitudes of the components vs. Temperature*

The simulated response of the normal force, and temperature difference for 1 vane-pair is presented in Figure 9, while Figure 10 presents the simulated shear force response.

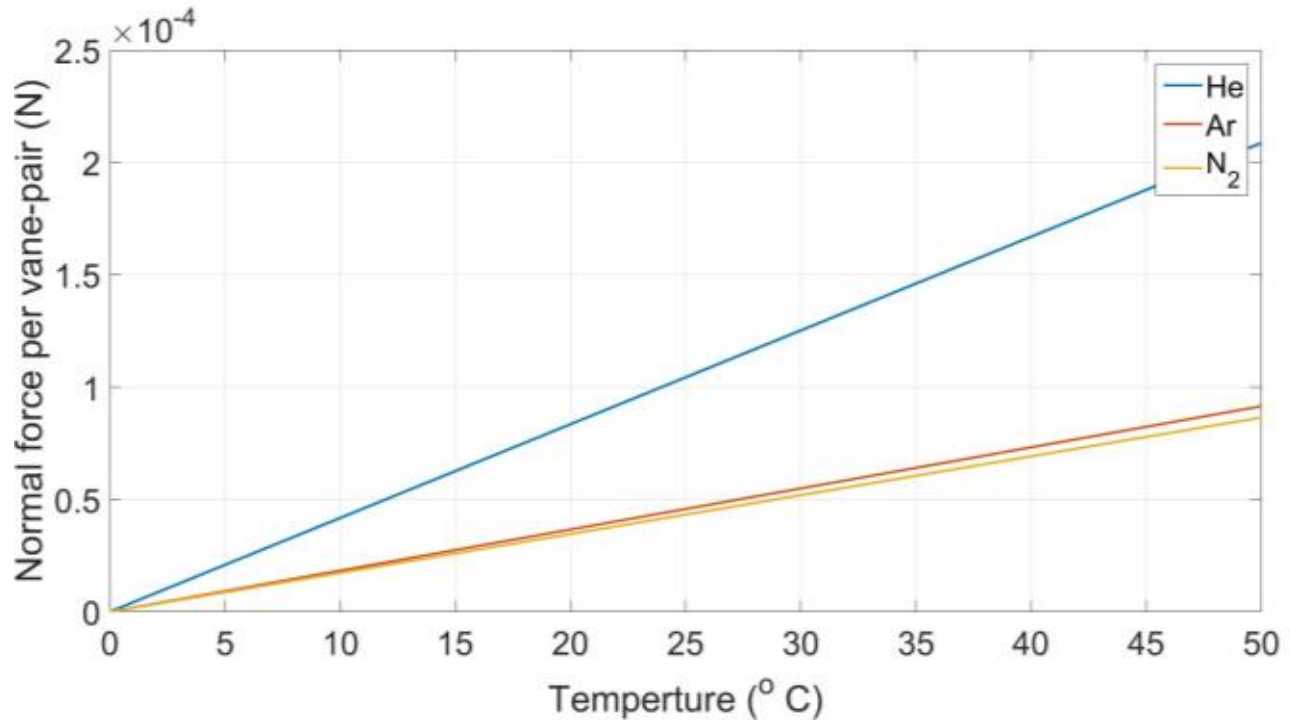

**Figure 9. Normal Force vs Temperature**

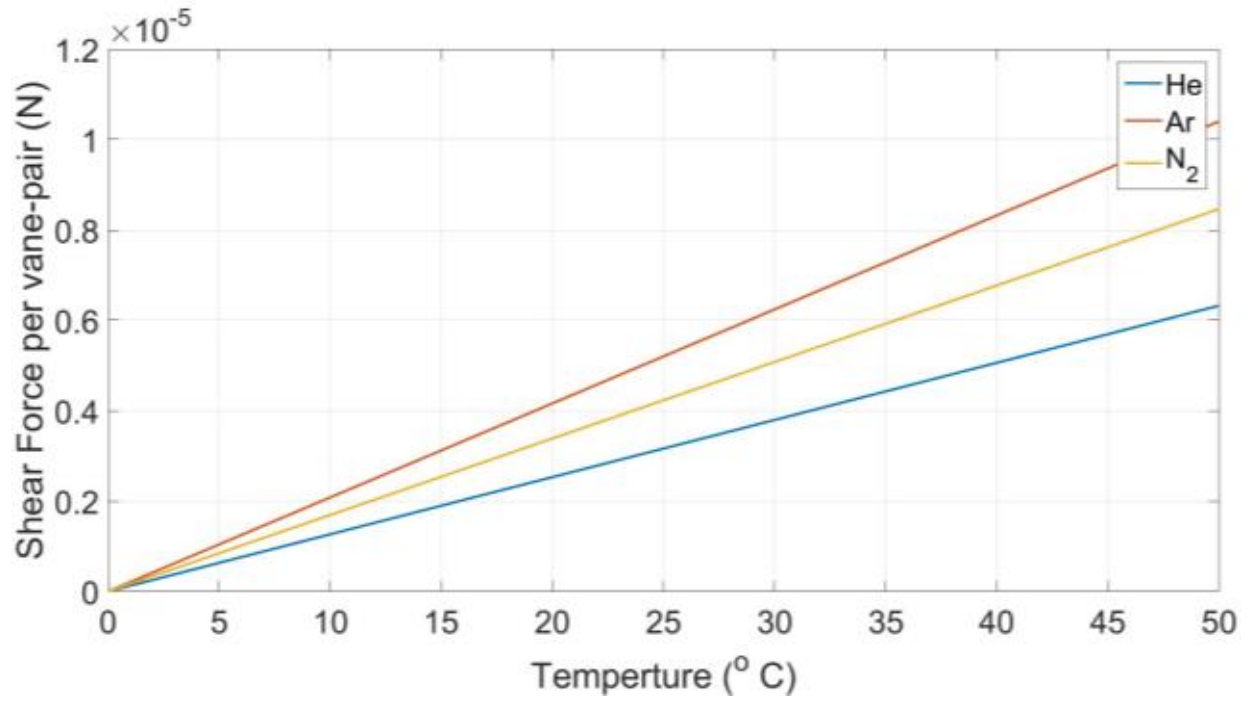

**Figure 10. Shear Force vs Temperature**

*Net actuator force vs. Temperature*

The net radiometric force generated per actuator is plotted as a function of the temperature difference between the vane pairs, as shown in Figure 11. As expected, the temperature dependence is linear for all the plots, since ΔT is a common factor for Equations 1, 8, and 12. For the 3U CubeSat proposed, if we assume a moment arm of 5 cm, a tangential force of about 42E-8 N (torque/moment arm) is generated by the solar radiation pressure. The temperature differences required by different gas is shown in Figure 11 and recorded in Table 1. Clearly, this proposed actuator, in theory, can generate enough torque to stabilize a spacecraft from solar radiation disturbance.

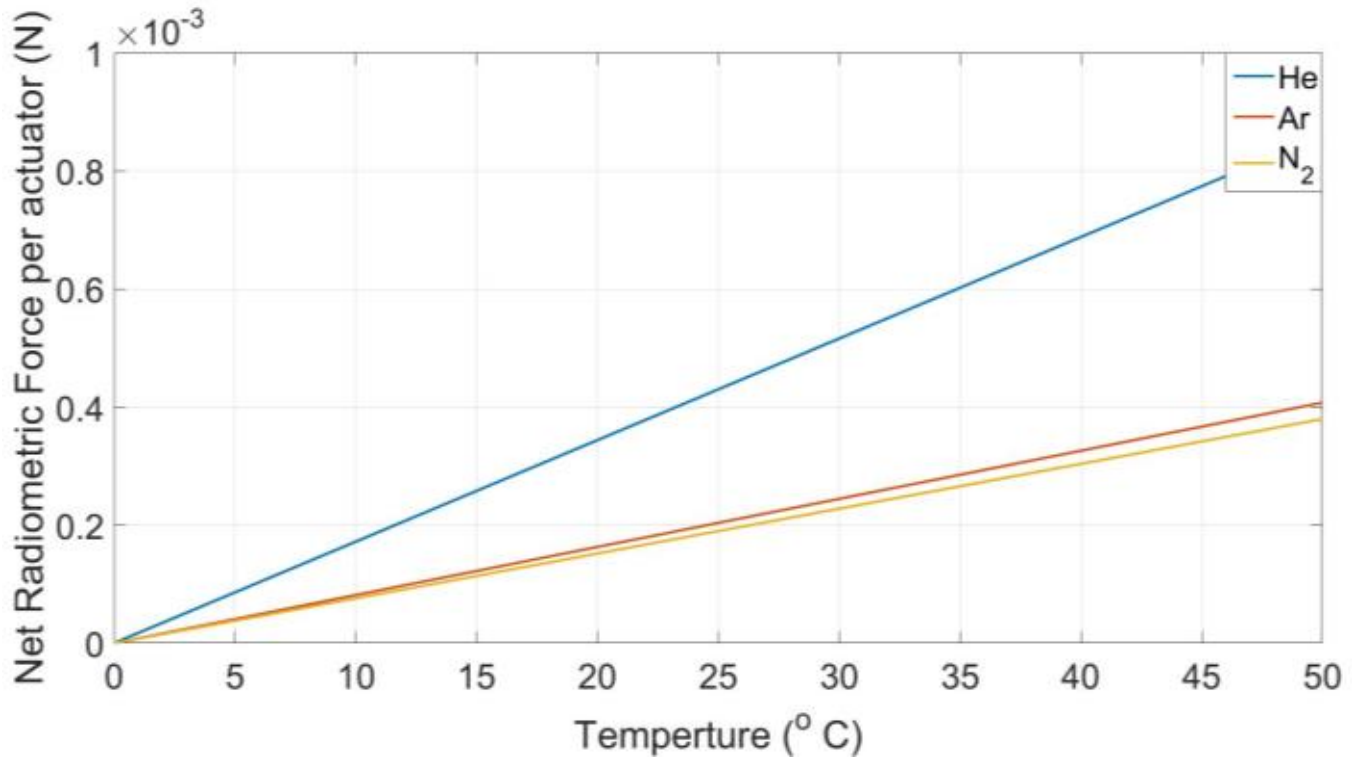

**Figure 11. Dependence of Temperature differences on generating the radiometric forces**

**Table 1. Gas Properties**

| Gas | σ (pm) | $P_{gas}$(Pa) | λ (nm) | α | ΔT (°C) |
|---|---|---|---|---|---|
| He | 260 | 13.52 | 389.1 | 0.51 | 0.024 |
| Ar | 348 | 7.56 | 217.2 | 0.83 | 0.051 |
| $N_2$ | 364 | 6.87 | 198.5 | 0.79 | 0.055 |

## 6. DISCUSSION

The proposed actuator design is shown to be suitable for attitude control in deep space. The actuator can compensate for natural disturbances from solar radiation pressure.

The design presented here is based on analytical formulations done by Scandurra, et al. [6] and this model has been known to be experimentally validated. However, Equation 8 is only valid for low to moderate temperature gradients [6]. Therefore, in order to model the actuator, the Boltzmann's transport equation needs to be solved. Computer packages solve them for a given model by using Direct Simulation Monte Carlo (DSMC) technique.

Presently, simulation studies of the actuator are being developed in SPARTA (Stochastic Parallel Rarefied gas Time Accurate solver), which is an open source DSMC tool developed by SANDIA laboratories. A comparison, of the analytical design proposed design could then be compared to its simulated performance. The steps succeeding this will be experimental validation with a physical prototype of the actuator.

## 7. CONCLUSION

This paper introduced a new class of attitude control actuators called radiometric actuators. These are suitable for CubeSat based telescope missions and for CubeSat centrifuges. Due to their low expected cost, this can substantially reduce the cost and reliability of the platforms. An analytical model of the radiometric forces is presented which is valid for low to medium temperature gradients and Knudsen numbers in the transition regime. This model was then used to describe the design procedure of the radiometric actuator. Some initial feasibility analysis of the forces generated by the device has been presented.

## ACKNOWLEDGEMENTS

The authors thank Dr. Sergey Gimelshein of University of Southern California for suggesting the use of SPARTA. The authors would also thank Dr. Steve Plimpton of Sandia Research Labs for helping us with SPARTA

## BIOGRAPHY

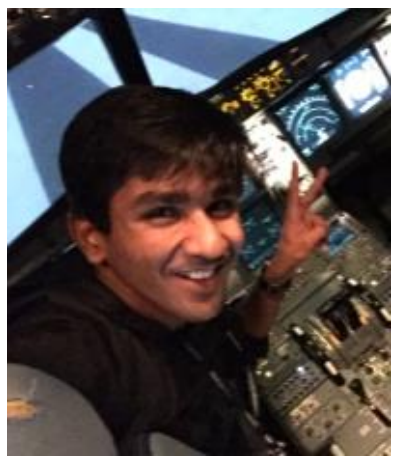

***Ravi Teja Nallapu*** *received a B.Tech. in Mechatronics Engineering from JNTU, Hyderabad, India in 2010. He then received an M.S in Aerospace Engineering from University of Houston, TX in 2012. After this, he worked with U.S Airways as a Flight Simulator Engineer for 2 years. He is presently pursuing his PhD in Aerospace Engineering from Arizona State University, Az. He specializes in control theory and robotics. His research interests include GNC of spacecrafts, space systems engineering, orbital mechanics, and exploration robotics.*

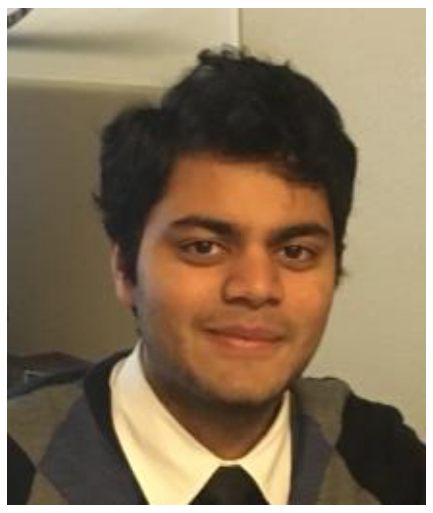

***Amit Tallapragada*** *is an undergraduate at Arizona State University pursuing a B.S in Computer Science. He is currently an undergraduate researcher at the Space and Terrestrial Robotic Exploration (SpaceTREx) Laboratory at Arizona State University. Here, he has primarily worked on the visualization of data and software development. His research interests are big data analytics, flight software, and artificial intelligence.*

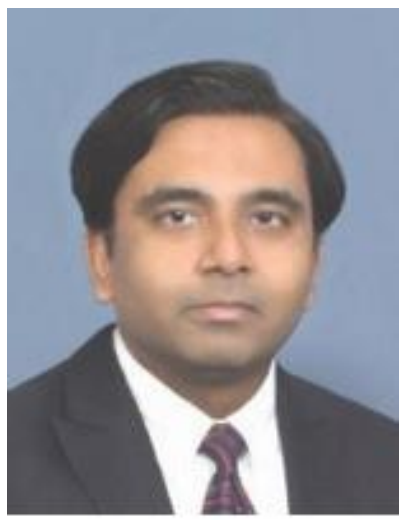

***Jekan Thangavelautham*** *is an Assistant Professor and has a background in aerospace engineering from the University of Toronto. He worked on Canadarm, Canadarm 2 and the DARPA Orbital Express missions at MDA Space Missions. Jekan obtained his Ph.D. in space robotics at the University of Toronto Institute for Aerospace Studies (UTIAS) and did his postdoctoral training at MIT's Field and Space Robotics Laboratory (FSRL). Jekan Thanga heads the Space and Terrestrial Robotic Exploration (SpaceTREx) Laboratory at Arizona State University. He is the Engineering Principal Investigator on the AOSAT I CubeSat Centrifuge mission and is a Co-Investigator on SWIMSat, an Airforce CubeSat mission to monitor space threats.*